\documentstyle[prc,aps,preprint]{revtex}
\begin{document}
\draft
\date{\today }
\title{Collisional Damping of Giant Monopole and Quadrupole Resonances}

\author{ S. Yildirim$^{1}$, A. Gokalp$^{1}$, O. Yilmaz$^{1}$ and S. Ayik$^{2}$}
\address{{\small $^{1}$ {\it Physics Department, Middle East Technical University,}\\
06531 Ankara, Turkey}}
\address{{\small $^{2}$ {\it Physics Department, Tennessee Technological University,}
Cookeville TN 38505, USA}}

\maketitle

\begin{abstract}
Collisional damping widths of giant monopole and quadrupole excitations
for $^{120}$Sn and $^{208}$Pb at zero and finite temperatures are calculated
within Thomas-Fermi approximation by employing the microscopic in-medium
cross-sections of Li and Machleidt and the phenomenological Skyrme and
Gogny forces, and are compared with each other. The results for
the collisional widths of giant monopole and quadrupole vibrations at zero
temperature as a function of the mass number show that the collisional
damping of giant monopole vibrations accounts for about $30-40\%$
of the observed widths at zero temperature, while for giant quadrupole
vibrations it accounts for only $20-30\%$ of the observed widths of
zero temperature.

\end{abstract}

~~~~~~~~~~~~\\

\pacs{ PACS:24.30.Cz, 24.30.Gd, 25.70.Ef, 25.70.Lm}

Nuclear collective vibrations built on the ground state and on the excited
states of the nucleus have been studied extensively during the last several
years both theoretically and experimentally \cite{R1}.
On the theoretical side, much effort have been devoted to understand the
damping properties of giant dipole excitations at zero and finite temperatures.
In medium-weight and heavy nuclei at relatively low temperatures the
damping results mostly from the spreading width $\Gamma^{\downarrow}$ which is
due to mixing of the collective state with the near by more complex doorway states
\cite{R2,R3,R4}. There are essentially two different approaches for calculation of
the spreading widths:
(i) Coherent mechanism due to coupling with low-lying surface modes which provides
an important mechanism for damping of giant resonance in particular at low
temperatures \cite{R5,R6,R7,R8},
(ii) Damping due to the coupling with incoherent 2p-2h states which is usually
referred to as the collisional damping \cite{R9,R10,R11,R12}, and the Landau damping modified
by two-body collisions \cite{R13,R14,R15}. The investigations carried out
on the basis of these approaches have been partially successful in explaining the
broadening of the giant dipole resonance with increasing temperature.
In this work, we do not consider the coherent contribution to damping,
but investigate the collisional damping of isoscalar giant monopole and isoscalar
giant quadrupole resonances at zero and finite temperatures due to decay of the
collective state into incoherent 2p-2h excitations in the basis of a semi-classical
non-Markovian
transport approach. In this approach, the collisional term involves two-body
transition rates which can be expressed in terms of the in-medium scattering
cross sections \cite{R10}. Therefore in order to assess the contribution of
collisional damping to the total width of giant resonance excitations, we need
realistic in-medium cross-sections which correctly interpolate between the
free space and the medium. In previous investigations, the contributions of
spreading width resulting from the decay of the collective state into incoherent
2p-2h states have been estimated by employing either free nucleon-nucleon
cross sections or an effective Skyrme force \cite{R9,R10}.
However, Skyrme force at most can provide a semiquantitative description of the
collisional damping since it provides a poor approximation in the collisional term
because in the vicinity of nuclear surface it does not match at all to the free
nucleon-nucleon cross-sections. 
In a recent work, we calculated the collisional damping width of the giant dipole
excitations by employing the microscopic in-medium cross sections of Li and 
Machleidt \cite{R16}, which interpolate correctly between the free space and 
the medium and provide the best available input for determining the magnitude
of the collisional damping \cite{R17}. 
In the present work, we extend this analysis to the study of the collisional damping
widths of giant monopole and quadrupole excitations by employing
the microscopic in-medium cross-sections of Li and Machleidt  and phenomenological
Skyrme and Gogny forces.

We study the collective vibrations in the small amplitude limit of the extended TDHF
theory in which damping resulting from the coupling of the collective state to
incoherent 2p-2h states is included in the form of a non-Markovian collision
term \cite{R18,R19}. In the Hartree-Fock representation, the Fourier transform
of the self-energy of  collective modes due to coupling with the incoherent 2p-2h states
is given by
\begin{equation}
\Sigma _{\lambda }(\omega )=\frac{1}{4}\sum \frac{\left|<ij|[O_{\lambda
}^{\dagger },v]|kl>_{A}\right|^{2}}{\hbar \omega -\Delta \epsilon +i\eta }[
n_{k}n_{l}\bar{n}_{i}\bar{n}_{j}-n_{i}n_{j}\bar{n}_{k}\bar{n}_{l}]
\end{equation}
where $O_{\lambda}^{\dagger }$ is the collective operator associated with the RPA
mode $\lambda$, $v$ is the effective interaction that couples the ph-space with
2p-2h configurations, $\bar{n}_{i}=1-n_{i}$,
$\Delta \epsilon =\epsilon _{i}+\epsilon_{j}-\epsilon_{k}-\epsilon _{l}$ ,
and $\eta $ is a small positive number \cite{R10}.
The real and imaginary parts of the self energy,
$\Sigma_{\lambda }(\omega )=
\Delta _{\lambda }(\omega )-\frac{i}{2}\Gamma_{\lambda}(\omega )$,
determine the energy shift and the damping width of the collective excitation,
respectively \cite{R3}.

We evaluate the expression for the self-energy in the Thomas-Fermi
approximation, which corresponds to the semi-classical transport description
of the collective vibrations. In Thomas-Fermi approximation
the self-energy of the collective modes can be deduced from the quantal 
expression  (1)
by replacing the occupation numbers $n_i$ with the equilibrium phase-space
density given by the Fermi-Dirac function as
$n_i \rightarrow f(\epsilon, T)=1/[exp(\epsilon-\mu)/T+1]$
with $\mu$ denoting the chemical potential, and summations over the 2p-2h states with
integrals over phase-space,
$\Sigma \rightarrow \int d{\bf r} d{\bf p}_1 d{\bf p}_2 d{\bf p}_3 d{\bf p}_4 $
 \cite{R10,R20}. Furthermore, spin-isospin effects in collective vibration can
be incorporated into the treatment by considering proton and neutron degrees of
freedom separately. Observing that in isoscalar modes protons and neutrons
vibrate in phase, in Thomas-Fermi approximation the collisional widths of
isoscalar modes can be expressed as \cite{R10}
\begin{equation}
\Gamma_{\lambda}^{s}=\frac{1}{N_{\lambda}}
\int d{\bf r} d{\bf p}_{1}d{\bf p}_{2}d{\bf p}_{3}d{\bf p}_{4} [W_{pp}+W_{nn}+ 2W_{pn}]
\frac {(\Delta \chi_{\lambda})^{2}}{2} Z f_{1}f_{2}\bar{f}_{3}\bar{f}_{4}
\end{equation}
where $N_{\lambda}=\int d{\bf r} d{\bf p}(\chi_{\lambda})^{2}
(-\frac{\partial}{\partial \epsilon}f)$
is a normalization,
$\Delta \chi_{\lambda}=\chi_{\lambda}(1)+\chi_{\lambda}(2)-
\chi_{\lambda}(3)-\chi_{\lambda}(4)$,
$Z= [\delta(\hbar\omega_{\lambda}-\Delta\epsilon)-
\delta(\hbar\omega_{\lambda}+\Delta\epsilon)]/ \hbar\omega_{\lambda}$, and
$\chi_{\lambda}(t)$ denotes the distortion factor of the phase-space density
$\delta f(t)= \chi_{\lambda}(t) (-\partial f/\partial \epsilon)$ in the
corresponding mode. In this expression, two-body transition rates
$W_{pp}, W_{nn}$ and $ W_{pn}$ associated with proton-proton, neutron-neutron and
proton-neutron collisions are given in terms of the corresponding scattering
cross-sections
\begin{equation}
W(12;34)= \frac{1}{(2\pi\hbar)^3}\frac{4\hbar}{m^2}\frac{d\sigma}{d\Omega}
\delta({\bf p}_{1}+{\bf p}_{2}-{\bf p}_{3}-{\bf p}_{4}).
\end{equation}

We apply the formula (2) to calculate the collisional widths of  monopole
and quadrupole vibrations. We use the nuclear fluidynamical model  to
express  the distortion factors of the momentum distribution in which the
distortion  factors can be expressed in terms of the velocity field
$\Phi ({\bf r})$ associated with the collective mode as
$\chi=({\bf p}\cdot {\bf \nabla})({\bf p}\cdot {\bf \nabla})\Phi ({\bf r})$. 
An accurate
description of the monopole vibrations can be obtained by parameterizing the
velocity field in terms of the zeroth order Bessel functions $\Phi (r)=j_0(kr)$,
with the wave number $ k=\pi/R $, and the nuclear radius R
\cite{R21,R22}. For the quadrupole vibrations, we use the parameterization
of the velocity field in terms of second order  Bessel function
$\Phi (r)=j_2(kr)$ and take  $k=3.34/R $ \cite{R23}.  We also carry out
the calculations for the collisional width of the quadrupole vibrations
by taking the distortion factor of the momentum distribution according to the
scaling picture as
$\chi_Q= p^2P_2(\cos\theta )$ \cite{R10}.
In the case of isoscalar modes, the collisional width is determined by the spin-isospin
averaged nucleon-nucleon cross-section,
$(d\sigma/d\Omega)_0=[(d\sigma/d\Omega)_{pp}+(d\sigma/d\Omega)_{nn}+
2(d\sigma/d\Omega)_{pn}]/4$.
The nucleon-nucleon cross-sections in this expression associated with an effective
residual interaction can be expressed as
\begin{equation}
\left(\frac{d\sigma}{d\Omega}\right)_{pp}=\left(\frac{d\sigma}{d\Omega}\right)_{nn}=
\frac{\pi}{(2\pi\hbar)^3}\frac{m^2}{4\hbar}\frac{1}{4}\sum_{S} (2S+1)
|<{\bf q};S,T=1|v |{\bf q}^{\prime};S,T=1>_{A}|^{2}
\end{equation}
and
\begin{equation}
\left(\frac{d\sigma}{d\Omega}\right)_{pn}=
\frac{\pi}{(2\pi\hbar)^3}\frac{m^2}{4\hbar}\frac{1}{8}\sum_{S,T} (2S+1)
|<{\bf q};S,T|v |{\bf q}^{\prime};S,T>_{A}|^{2}~~,
\end{equation}
where ${\bf q}=({\bf p}_{1}-{\bf p}_{2})/2$,
${\bf q}^{\prime}=({\bf p}_{3}-{\bf p}_{4})/2$ are the relative momenta before and
after a binary collision, and $<{\bf q};S,T|v|{\bf q}^{\prime};S,T>_{A}$ represents the
fully anti-symmetric matrix element of the residual interaction between two
particle states with total spin and isospin $S$ and $T$.
By noting that, S=T=1 and S=T=0 matrix elements of the interaction are space
antisymmetric, and S=1, T=0 and S=0, T=1 matrix elements are space
symmetric, we find that the spin-isospin averaged nucleon-nucleon cross-section
associated with the Gogny force is given by
\begin{eqnarray}
\left( \frac{d\sigma}{d\Omega}\right)_{0}^G & = &
\frac{\pi}{(2\pi\hbar)^3}\frac{{m_G^*}^2}{4\hbar}\frac{1}{8}\nonumber \\
&\times& \left\{ \frac{9}{2} \left| \sum_{i=1}^2 I_i^{-}(W_i + B_i - H_i - M_i ) \right|^2
+ \frac{1}{2}\left| \sum_{i=1}^2 I_i^{-}(W_i -B_i + H_i -M_i ) \right|^2+ \right.  \nonumber \\
& & \left. \frac{3}{2} \left| \sum_{i=1}^2 I_i^{+}( W_i + B_i + H_i + M_i) +4 t_3
\rho^{1/3}\right|^2+ \frac{3}{2}\left| \sum_{i=1}^2 I_i^{+}(W_i - B_i - H_i + M_i
)\right|^2 \right \}
\end{eqnarray}
where $m_G^{*}$ denotes the effective mass corresponding to the Gogny force,
and the quantities $I_i^{+}$ and $I_i^{-}$ are the symmetric and
anti-symmetric matrix elements of the Gaussian factor in the force,
\begin{equation}
I_{i}^{\pm}= (\sqrt{\pi}\mu_i)^3 \left( exp [- \frac{1}{4} ({\bf q}-{\bf q}%
^{\prime})^2 (\frac {\mu_i}{\hbar})^2] \pm exp [- \frac{1}{4} ({\bf q}+ {\bf %
q}^{\prime})^2 (\frac {\mu_i}{\hbar})^2] \right).
\end{equation}
In these expressions, $\rho$ is the local density and $W_i , B_i , H_i ,
M_i, \mu_i $ denote the standart parameters of the Gogny force \cite{R24,R25}.
In a similar manner, this cross-section can be calculated in terms of the effective
Skyrme force as \cite{R10,R17}
\begin{eqnarray}
\left( \frac{d\sigma}{d\Omega}\right)_{0}^S =
\frac{\pi}{(2\pi\hbar)^3}\frac{{m_S^*}^2}{4\hbar}
& &\left(\frac{3}{4} [t_0 (1-x_0 )+\frac{t_1}{
2\hbar^2} (1-x_1)({\bf q}^2+{{\bf q}^\prime}^2)+ \frac{t_3}{6}
(1-x_3)\rho^{\alpha}]^2+ \right. \nonumber \\
& & \left. \frac{3}{4}[t_0 (1+x_0 )+ \frac{t_1}{2\hbar^2}(1+x_1)({\bf q}^2+{{\bf q}
^\prime}^2)+ \frac{t_3}{6}(1+x_3) \rho^{\alpha}]^2 + \right.  \nonumber \\
& & \left. \frac{5}{8}[ (\frac {t_2}{\hbar^2})^2 (1-x_2)^2 + 3 (\frac {t_2}{\hbar^2})^2
(1+x_2)^2] ({\bf q} \cdot {\bf q}^\prime)^2 \right)~~.
\end{eqnarray}
The expressions for the effective masses $m_G^*( r )$ and $m_S^*( r )$ for
the Gogny and Skyrme forces are given in \cite{R24,R25}.
The angle $\Theta$ between ${\bf q}$ and ${\bf q}^{\prime}$ in the
cross-sections (6) and (8) defines the scattering angle in the center of
mass frame, and the total cross-section is found by an integration over this
angle, $\sigma_{0}=2\pi \int sin\Theta d\Theta \left( d\sigma/d\Omega
\right)_{0}$.
Microscopic in-medium cross-sections of Li and Machleidt \cite{R16}
for proton-proton and neutron-proton cross-sections are parameterized as
\begin{equation}
\sigma_{pp}^{LM}= [23.5+0.00256 \left (18.2-E_{lab}^{0.5}\right)^{4.0}] \;
\frac{1.0+0.1667 \; E_{lab}^{1.05}\; \rho^3} {1.0+9.704  \; \rho^{1.2}}~~,
\end{equation}
\begin{equation}
\sigma_{pn}^{LM}= [31.5+0.092 \left|20.2-E_{lab}^{0.53}\right|^{2.9}] \;
\frac{1.0+0.0034 \; E_{lab}^{1.51}\; \rho^2} {1.0+21.55  \; \rho^{1.34}}~~,
\end{equation}
where $E_{lab}=({\bf p}_1-{\bf p}_2)^2/2m=2 {\bf q}^2/m$ is the kinetic
energy of the projectile in the rest frame of the target nucleon which is
also equal to twice the energy available in the center of mass frame.

In figure 1 and  figure 2, we compare the spin-isospin averaged nucleon-nucleon
cross-sections calculated using the microscopic in-medium cross-sections of Li and
Machleidt (dotted lines) with the cross-sections of the Gogny force  (dashed lines)
and the Skyrme force (solid lines) with SkM$^*$ parameters.
In figure 1, the total cross-sections are plotted as a function of the projectile
energy $E_{lab}$ at two different nuclear matter densities
$\rho=\rho_{0}=0.18~~ fm^{-3}$ (top panel) and $\rho=\rho_{0}/2$ (bottom
panel). The cross-sections shown in the left and right parts of the figure
are calculated with the bare nucleon mass and the corresponding effective
masses, respectively. Around the normal nuclear matter density $\rho \approx \rho_{0}$,
and over a narrow energy interval around 150 MeV, these cross-sections roughly match,
however at lower densities and at lower and higher energies the phenomenological
cross-sections deviate strongly from the microscopic cross-sections,
deviations being larger in the calculations with the bare nucleon mass.
In figure 2, the cross-sections are shown as a function of density for the bombarding
energy  $E_{lab}=100$ MeV. The microscopic calculations approach the free
nucleon-nucleon cross-section for decreasing density, on the other hand
the phenomenological cross-sections strongly increase and reach large values
in free space. Therefore, we can safely state that the microscopic calculations of
Li and Machleidt provide a more reliable description of the in-medium cross-sections
than those given by the finite range Gogny and the zero range Skyrme force.
In a previous work \cite{R17}, the momentum integrals in the expression for the
damping width of giant dipole resonance were evaluated exactly. We follow the same
method in the present work and evaluate the momentum integrals in the expression (2)
exactly for monopole and quadrupole vibrations. In this calculation the angular
anisotropy of the cross-sections are neglected and we make the replacement
$\left( d\sigma/d\Omega \right)_{0}\rightarrow \sigma_0/4\pi$.
In the numerical evaluations, we determine the nuclear density $\rho( r )$
in Thomas-Fermi approximation using a Wood-Saxon potential with a depth $V_0=-44$ MeV,
thickness $a=0.67$ fm and sharp radius $R_0= 1.27 A^{1/3}$ fm, and we calculate the
position dependent chemical potential $\mu(r,T)$ in the Fermi-Dirac function
$f(\epsilon, T)$ at each temperature. We use the formula $\hbar \omega=64 A^{-1/3}$ MeV
for the giant quadrupole resonance energies, and the expressions
$\hbar \omega=31.2 A^{-1/3}+20.6 A^{-1/6}$ MeV for $A\geq 70$ and
$\hbar \omega=17.5$ MeV for $A <70$ to calculate the giant monopole resonance energies.
Figure 3 shows the collisional damping width of giant monopole resonance in $^{120}Sn $
and $^{208}Pb $ as a function of temperature.
In figure 4, we show the collisional damping width of giant quadrupole resonance in
$^{120}Sn $ and $^{208}Pb $ as a function of temperature, where we use the distortion
factors of momentum distributions obtained from the  fluidynamical model by
the parametrization $\Phi (r)=j_2(kr)$ of the velocity field. The collisional damping
widths of the same nuclei calculated by employing the distortion factor
$\chi_Q= p^2P_2(\cos\theta )$ using scaling approximation are shown in figure 5.
In these figures  T=0 experimental data points are also indicated.
Calculations performed with the cross-sections of Li and Machleidt are denoted
with dotted lines. For comparison, we also show the  results with the SkM$^*$
(solid lines) and the Gogny (dashed lines) cross-sections with the bare nucleon mass
(lower panel) and the effective nucleon mass (upper panel). For both monopole and
quadrupole vibrations, the calculations with cross-sections of Li and Machleidt
exhibit a weaker temperature dependence than the calculations with the SkM$^*$
and the Gogny  cross-sections with the bare nucleon mass and result in
considerably smaller damping widths at all temperatures than those with the
cross-sections of the phonemenological forces. However, if  the effective nucleon mass is used,
the calculations employing microscopic and phenomenological cross-sections give
collisional damping widths that almost agree with each other at low temperatures with
differences becoming somewhat more appreciable only at high temperatures.
Furthermore, a comparison of  figure 4 and figure 5 shows that the calculations
for quadrupole vibrations employing the scaling approximation for the distortion
factors result in smaller values for the damping width as a function of temperature
than those where the distortion factors are determined from fluidynamical picture.
Moreover, in calculations using the effective nucleon mass,
scaling approximation for the distortion factors give results for the collisional
damping widths that become somewhat more appreciably different for microscopic
and phenomenological cross-sections at relatively high temperatures.
In figure 6, we show the collisional damping widths of giant monopole (left) and
quadrupole vibrations (right) as a function of the mass number at zero temperature
calculated with the microscopic cross-sections (dotted lines), and
with  the SkM$^*$ (solid lines) and the Gogny (dashed lines) cross-sections. 
Here, we use the effective nucleon mass and employ the scaling approximation
for distortion factors in quadrupole vibrations,  since at zero temperature both 
scaling and fluidynamical pictures for distortion factors give results that are not
appreciably different from each other.
If we base our conclusion on the result obtained employing the microscopic
in-medium cross-sections,  we can conclude that the collisional damping of giant
monopole
vibrations accounts for about $30-40\%$ of the observed widths at zero temperature,
while for giant quadrupole vibrations it accounts for only $20-30\%$
of the observed widths at zero temperature.

\begin{center}

{\bf Acknowledgments}

\end{center}

One of us (S. A.) gratefully acknowledges the Physics Department of Middle
East Technical University for warm hospitality extended to him during his
visits. We thank R. Machleidt for providing a table of their cross-sections,
and P. Shuck for fruitful discussions.
This work is supported in part by the U.S. DOE Grant No. DE-FG05-89ER40530.


\newpage

{\bf Figure Captions:}

\begin{description}
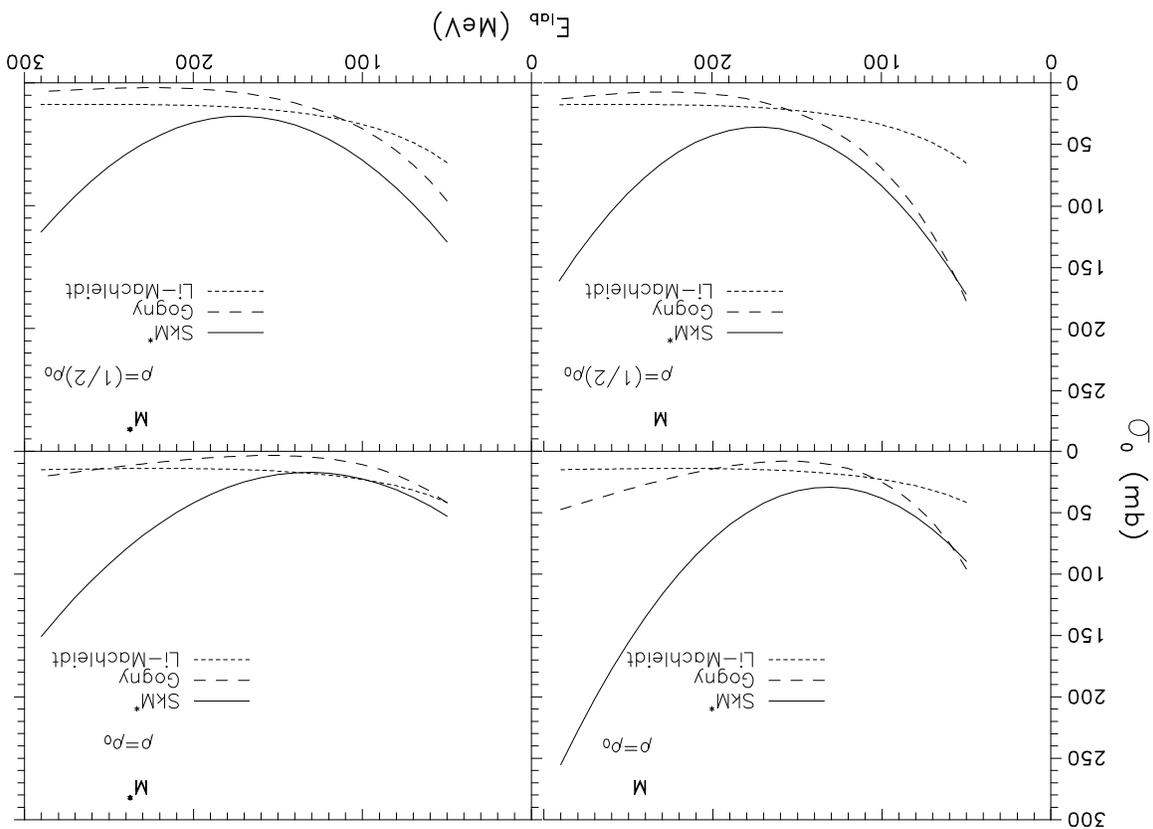
\item[{\bf Figure 1}:]  The spin-isospin averaged nucleon-nucleon
in-medium cross-sections as a function of bombarding energy $E_{lab}$
at several different densities.
Dotted lines are cross-sections of Li and Machleidt, and solid and dashed
lines are cross-sections associated with the SkM$^{*}$ and the Gogny forces
with the bare nucleon mass (left) and the effective nucleon mass (right),
respectively.

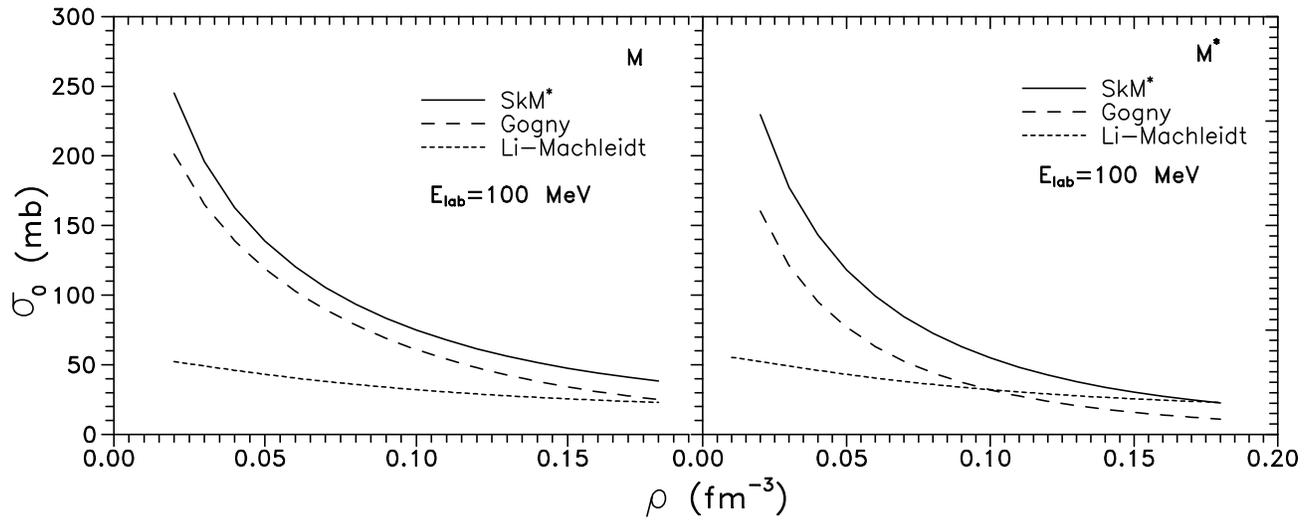
\item[{\bf Figure 2}:]  The spin-isospin averaged nucleon-nucleon
 in-medium cross-sections as a
function of density $\rho $ at $E_{lab}=100MeV$. Dotted lines are
cross-sections of Li and Machleidt, and solid and dashed lines are
cross-sections with the SkM$^{*}$ and the Gogny forces with the bare nucleon
mass (left) and the effective nucleon mass (right), respectively.

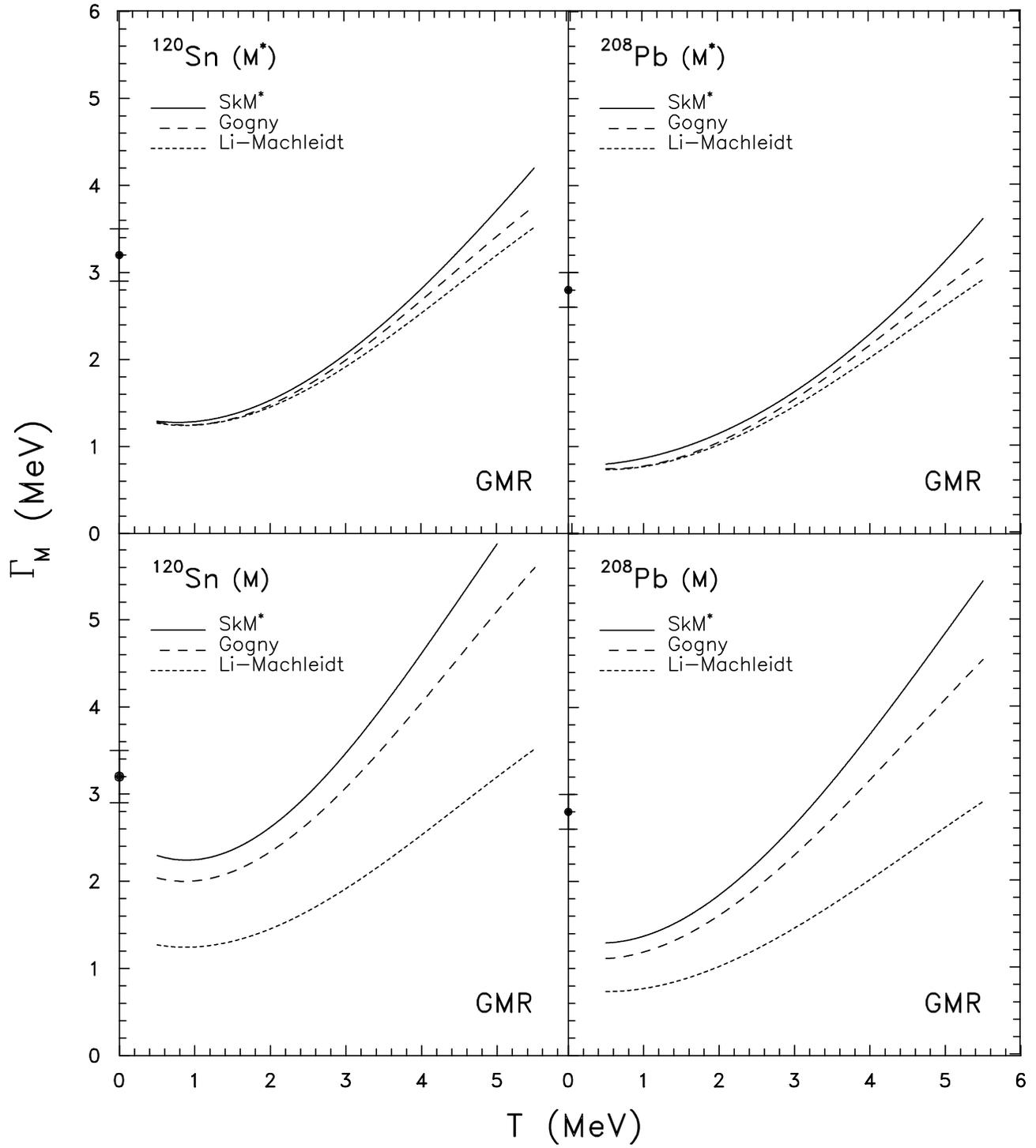
\item[{\bf Figure 3}:]  The collisional damping width of giant monopole resonance
 in $^{120}Sn$ and $^{208}Pb$ as a function of temperature.
 Dotted lines are calculations
with the cross-sections of Li and Machleidt, and solid and dashed lines are
results with the SkM$^{*}$ and the Gogny cross-sections with the bare
nucleon mass (lower panel) and the effective nucleon mass (upper panel).

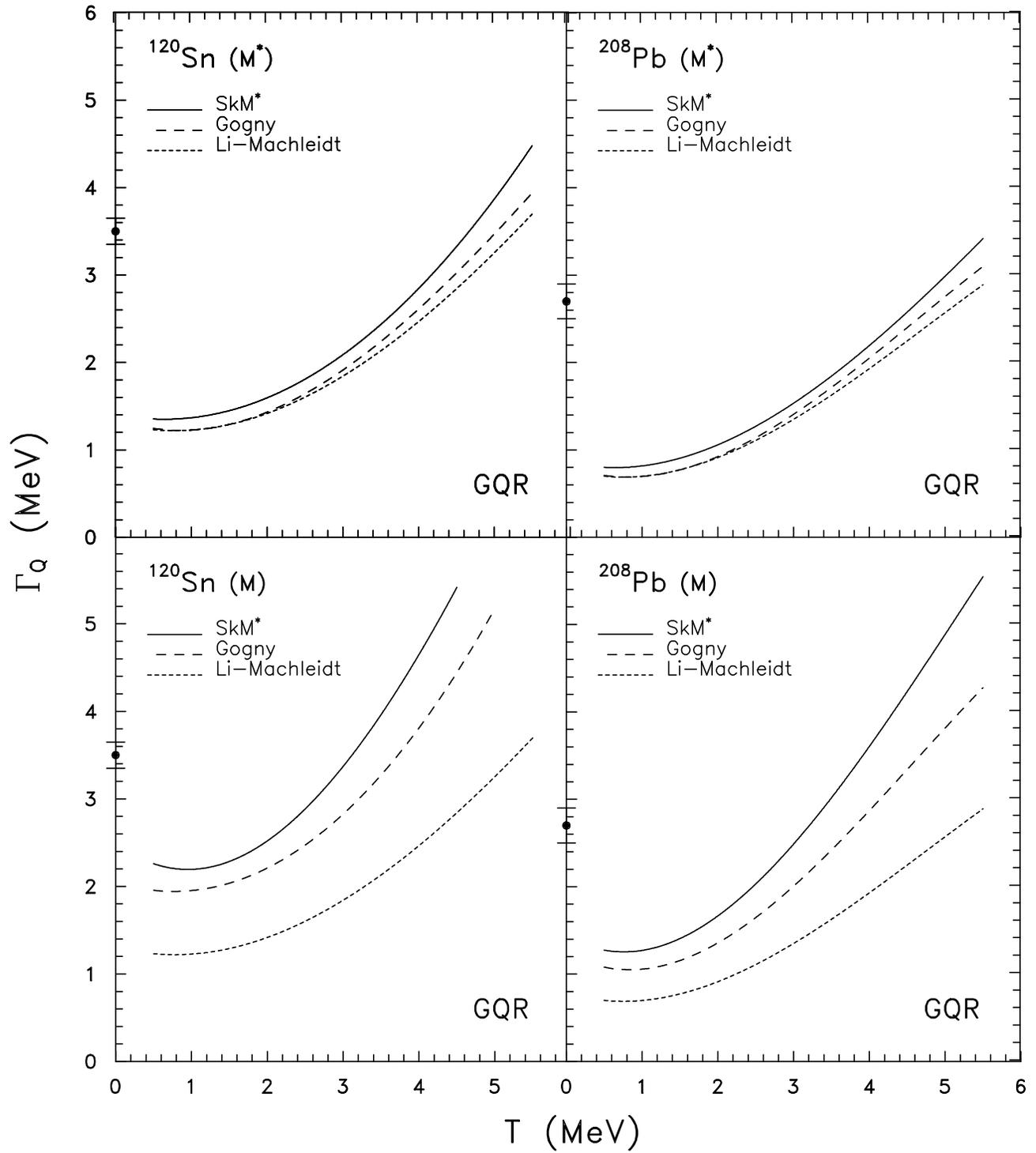
\item[{\bf Figure 4}:]   The collisional damping width of giant quadrupole
 resonance  in $^{120}Sn$ and $^{208}Pb$ as a function of temperature.
 Dotted lines are calculations with the cross-sections of Li and Machleidt,
 and solid and dashed lines are results with the SkM$^{*}$ and the Gogny
 cross-sections with the bare nucleon mass (lower panel) and the effective
 nucleon mass (upper panel). For distortion factors fluidynamical model
 is used.

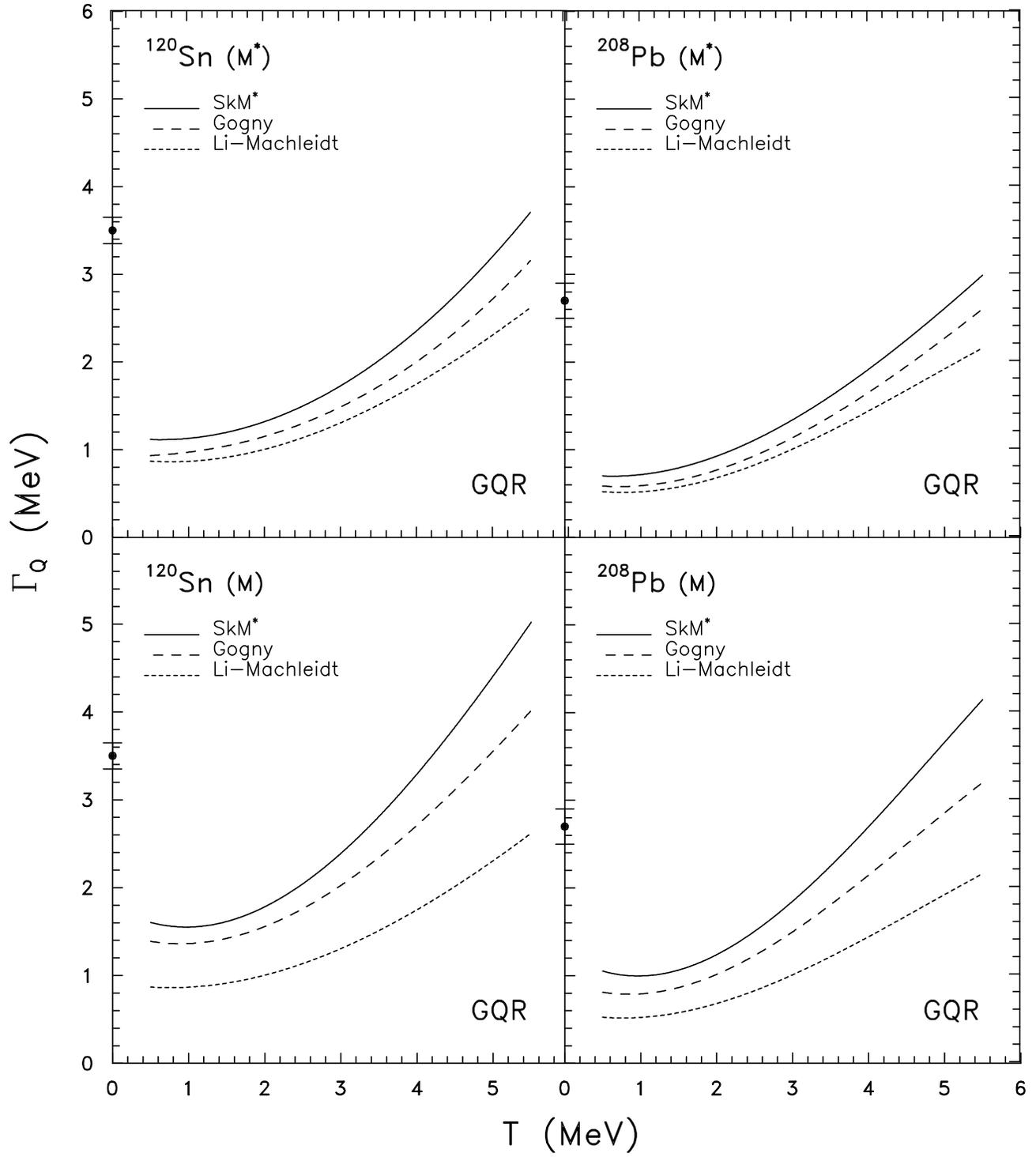
\item[{\bf Figure 5}:]   The collisional damping width of giant quadrupole
 resonance  in $^{120}Sn$ and $^{208}Pb$ as a function of temperature.
 Dotted lines are calculations with the cross-sections of Li and Machleidt,
 and solid and dashed lines are results with the SkM$^{*}$ and the Gogny
 cross-sections with the bare nucleon mass (lower panel) and the effective
 nucleon mass (upper panel). For distortion factors scaling approximation
 is used.

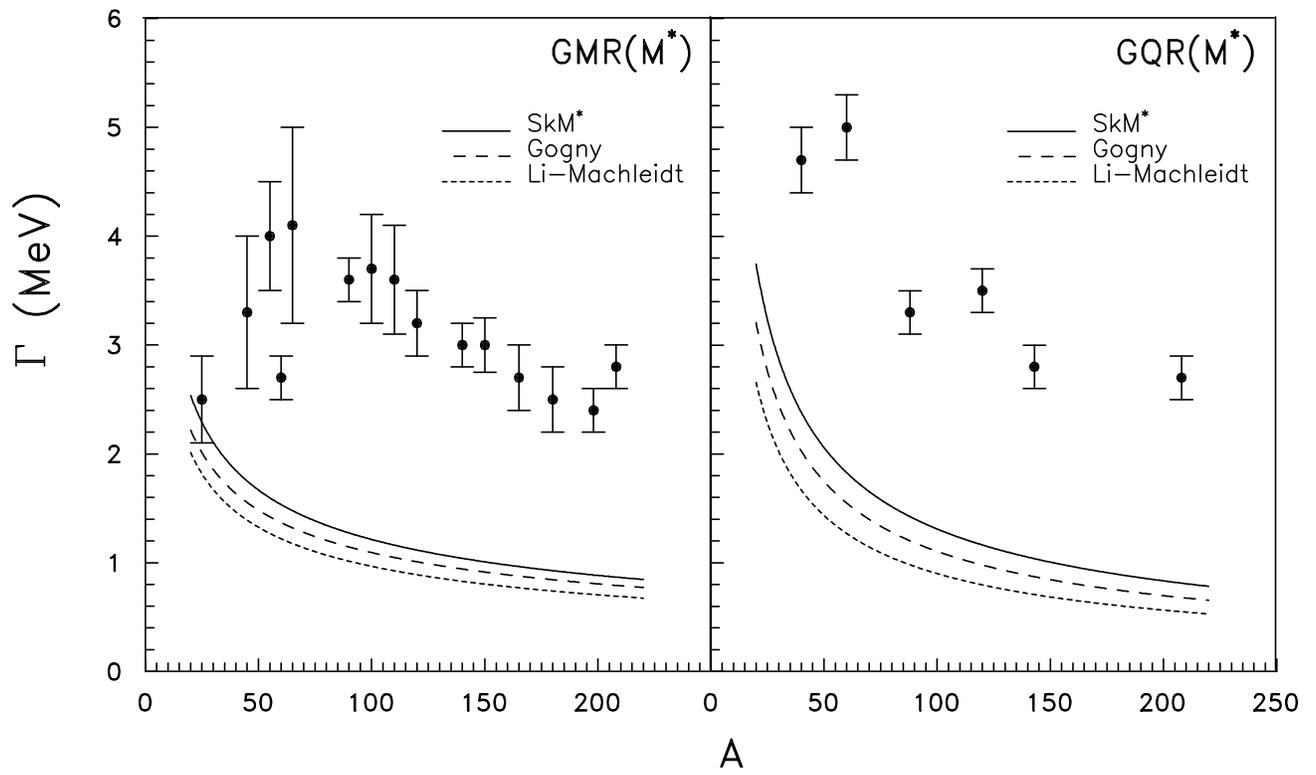
\item[{\bf Figure 6}:] The collisional widths of giant monopole (left) and
giant quadrupole (right) vibrations as a function of the mass number at
zero temperature with effective nucleon mass. Experimental widths are shown
by solid dots with error bars.

\end{description}

\end{document}